\documentclass{aa}  

\usepackage{graphicx}
\usepackage{txfonts}
\usepackage[hidelinks]{hyperref}
\hypersetup{colorlinks=true,linkcolor=blue,citecolor=blue,filecolor=blue,urlcolor=blue,}
\usepackage[dvipsnames]{xcolor}
\usepackage{subfigure}
\usepackage{newtxtext,newtxmath}
\usepackage{amsmath, amssymb}
\usepackage{mathtools}
\usepackage[T1]{fontenc}
\DeclareRobustCommand{\VAN}[3]{#2}
\let\VANthebibliography\thebibliography
\def\thebibliography{\DeclareRobustCommand{\VAN}[3]{##3}\VANthebibliography}
\usepackage{graphicx}	% Including figure files
\usepackage{amsmath}	% Advanced maths commands
\usepackage{booktabs}
\usepackage{subfigure}
\usepackage{float}

\begin{document}

   \title{Individual chaotic behaviour of the S-stars in the Galactic centre}
   
   \author{Sam J. Beckers\thanks{E-mail: beckers@strw.leidenuniv.nl}\inst{\ref{inst1}},
          Colin M. Poppelaars\thanks{E-mail: cpoppelaars@strw.leidenuniv.nl}\inst{\ref{inst1}},
          Veronica S. Ulibarrena\inst{\ref{inst1}}, Tjarda C.N. Boekholt\inst{\ref{inst2}} and Simon F. Portegies Zwart\inst{\ref{inst1}}
          }

   \institute{Leiden Observatory, Leiden University, 2300 RA, Leiden, The Netherlands \label{inst1}\and NASA Ames Research Center, Moffett Field, CA 94035, USA\label{inst2}}

   \date{Received ...; accepted ...}

% \abstract{}{}{}{}{} 
% 5 {} token are mandatory
 
  \abstract
    {Located at the core of the Galactic Centre, the S-star cluster serves as a remarkable illustration of chaos in dynamical systems. The long-term chaotic behaviour of this system can be studied with gravitational $N$-body simulations. By applying a small perturbation to the initial position of star S5, we can compare the evolution of this system to its unperturbed evolution. This results in the two solutions diverging exponentially, defined by the separation in position space $\delta_{r}$, with an average  Lyapunov timescale of $\sim$420 yr, corresponding to the largest positive Lyapunov exponent.  Even though the general trend of the chaotic evolution is governed in part by the supermassive black hole Sagittarius $\rm A^{*}$ (Sgr $\rm A^{*}$), individual differences between the stars can be noted in the behaviour of their phase-space curves.
    
    We present an analysis of the individual behaviour of the stars in this  Newtonian chaotic dynamical system. The individuality of their behaviour is evident from offsets in the position space separation curves of the S-stars and the black hole. We propose that the offsets originate from the initial orbital elements of the S-stars, where Sgr $\rm A^{*}$ is considered in one of the focal points of the Keplerian orbits. Methods were considered to find a relationship between these elements and the separation in position space. Symbolic regression turns out to provide the clearest diagnostics for finding an interpretable expression for the problem. Our symbolic regression model indicates that $\left\langle\delta_r\right\rangle \propto e^{2.3}$, implying that the time-averaged individual separation in position space increases rapidly with the initial eccentricity of the S-stars.}


   \authorrunning{S.J. Beckers et al.}
   \titlerunning{Individual chaotic behaviour of the S-stars}
   \maketitle
%
%-------------------------------------------------------------------
\section{Introduction}

    In the pursuit of understanding the nature of chaos and its manifestation in the universe, the S-star cluster is used as a laboratory for experimenting with the underlying astrophysics that give rise to chaos. The orbital data of 27 S-stars orbiting the supermassive black hole Sagittarius $\rm A^{*}$ from \citep{Gillessen_2009} are employed as initial conditions for a Newtonian gravitational $N$-body simulation of the S-star cluster. The orbital evolution of the system was tracked over $10^4$ yr by \citep{portegies-zwart-2023, 2023IJMPD..2342003B}. When perturbing the initial conditions of the star S5 by displacing its initial $x$ coordinate by $dx = 15$m, the evolution of the perturbed solution diverges exponentially from the canonical solution, implying that the system is chaotic. 
    
    In this paper, we present the methods used to quantify chaos in the simulation of the S-star cluster and delve into the individual differences in the chaotic behaviour of the S-stars and Sgr A*. For more details on the simulation, refer to \citep{portegies-zwart-2023}.

%--------------------------------------------------------------------
\section{Measuring chaos}

    In the gravitational $N$-body problem, chaos can be measured with the Lyapunov timescale. The Lyapunov timescale represents the timescale on which the system becomes unpredictable. We describe the chaotic S-star orbital evolution with phase-space distance. Phase-space distance as a function of time is defined as
    \begin{equation}
     \delta^2 = \frac{1}{4}\left(|\vec{r}_{c} - \vec{r}_{p}|^2 + |\vec{v}_{c} - \vec{v}_{p}|^2\right) = \frac{1}{4}\left(\delta_{r}^2 + \delta_{v}^2\right),
    \end{equation} 
    where $\vec{r}$ and $\vec{v}$ are the position and velocity vectors of an S-star, and $c$ and $p$ denote the canonical and perturbed solutions, respectively. We define an (initial) perturbation at $t=0$, $\delta(0)$. The evolution of $\delta(t)$ is then approximately  described by an exponential function with time dependency, 
    \begin{equation}\label{eq:lyaexp}
        \delta(t)=\delta(0)e^{\lambda t},
    \end{equation}
    where $\lambda$ is the maximum positive Lyapunov exponent. The growth factor, $G_{\delta}(t)$, is the value of $\delta$ at some time $t$ as a fraction of the initial perturbation, $\delta(0)$, i.e. 
    \begin{equation}
        G_{\delta}\equiv \frac{\delta(t)}{\delta(0)}=e^{\lambda t}.
    \end{equation}
    From this equation, one can find $\lambda=\rm ln(G_{\delta})/t$, which is the reciprocal of the Lyapunov timescale,
    \begin{equation}
        t_{\lambda}=\frac{1}{\lambda}=\frac{t}{\rm ln(G_{\delta})}.
    \end{equation}
    
    \subsection{Separation in position space}\label{sec:separation}
\begin{figure*}[h]
        \centering \includegraphics[width=0.9\linewidth]{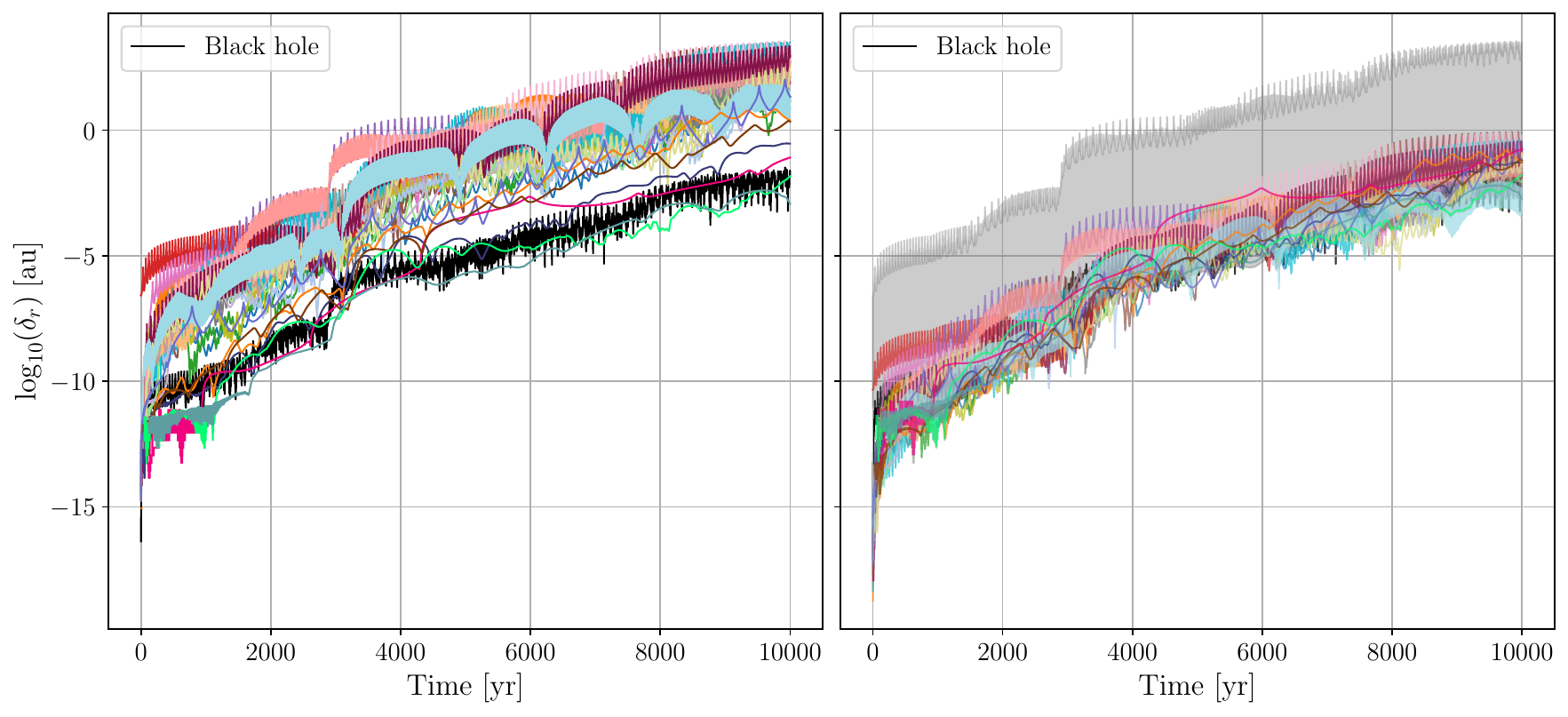}
        \caption{(left) Time evolution of separation in position space for each S-star and the central black hole. (right) $y = - 0.00018446 \cdot a/\mathrm{au} + \log(e) + 4.3103$ subtracted from the time evolution of $\log_{10}(\delta_r)$ for each S-star. The grey shading shows the region between the non-reduced maxima and minima of $\delta_r$ over the whole system at each timestep, indicating the magnitude of reduction in the spread of the curves,  though its lower section is obscured by the coloured curves. }
        \label{fig:pysrfinal}
    \end{figure*}

    We calculated the time evolution of the separation in position space between the canonical and perturbed solution, $\delta_r$, for each S-star and Sgr A*. We show this time evolution in the left panel of \autoref{fig:pysrfinal}.
    The separation in position space grows approximately exponentially, as expected from \autoref{eq:lyaexp}. From a least-squares fit to the curve of each star and the black hole, we infer an ensemble mean Lyapunov timescale of $t_{\lambda} \simeq 420$ yr. The black hole is less sensitive to the perturbation than most stars, as seen from its $\delta_r$ values, which are generally lower in magnitude than those of the S-stars. We attribute this to the mass difference of order $10^5 \; \rm{M}_{\odot}$ between the central body and the stars, leading to much higher inertia. This stabilizes the black hole's position, even in the presence of perturbations. Perturbations caused by close encounters between stars are propagated through the entire system, driven by feedback from the black hole \citep{portegies-zwart-2023}. Hence, big events with multiple close encounters, such as the one at $t = 2876$ yr, influence the general trend of all curves. However, it is not immediately evident why there are vertical offsets in the S-star curves, as the S-stars have been given identical masses in this simulation. Furthermore, while differences in the shapes of the individual curves could account for an offset, the curves as a whole exhibit a noticeable shift relative to each other.
%-----------------------------------------------------------------
\section{Position space separation offsets}
    The systematic offsets found in the position space separation curves must be an effect of some underlying astrophysics. The (initial) S-star orbits can introduce variations in the evolution of $\delta_r$ of each star, so it is worthwhile to investigate the Keplerian elements.
    
    \subsection{A relationship between the offsets in position space separation and Keplerian elements}
\begin{figure*}[h]
        \vspace{-0.375cm}
        \centering \includegraphics[width=16cm]{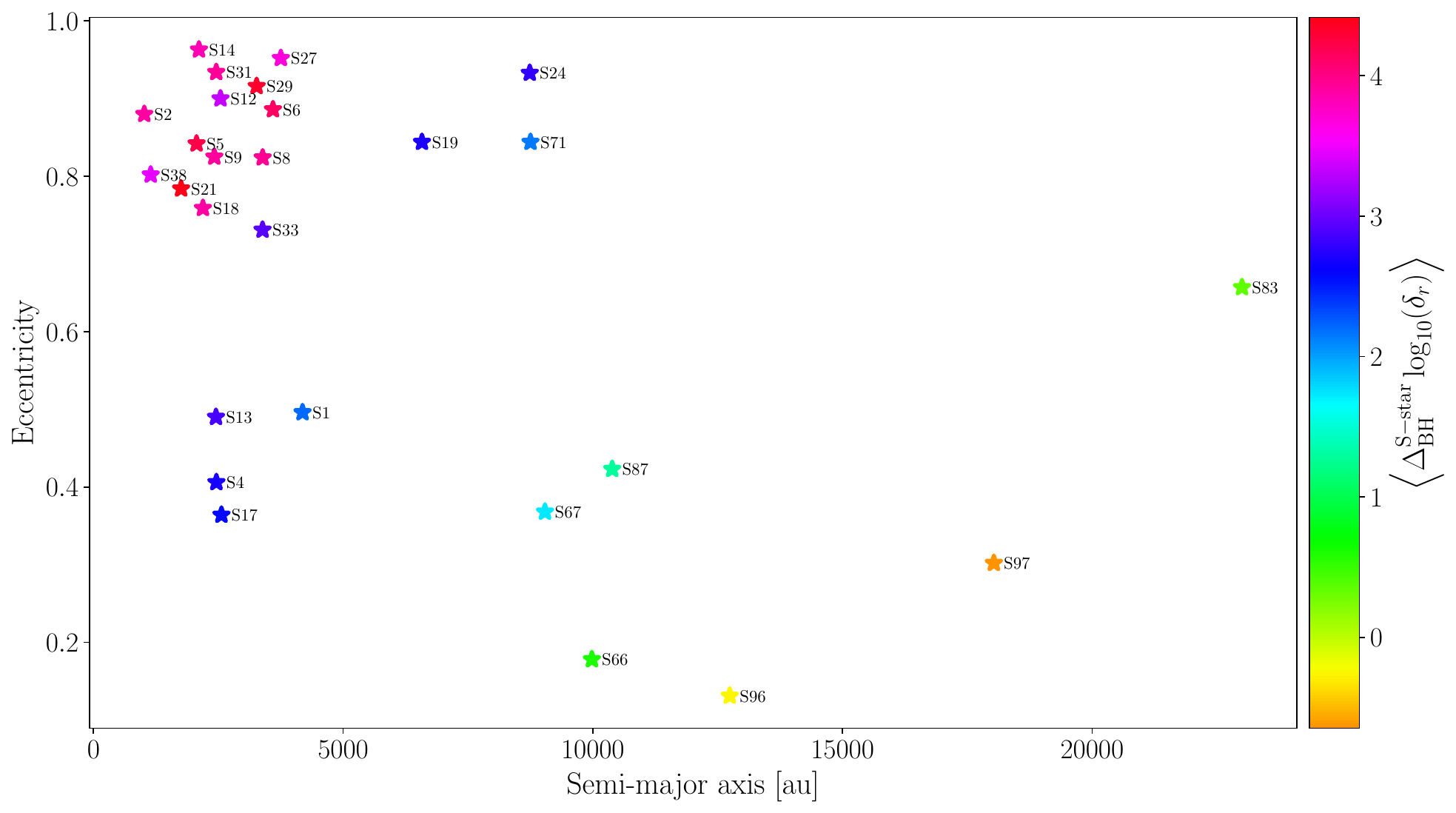}
        \caption{Initial semi-major axis and eccentricity of each S-star, with $\left\langle \Delta^{\mathrm{S\text{-}star\vphantom{{}_\mathrm{BH}}}}_{\mathrm{BH}}\log_{10}(\delta_r)\right\rangle$ indicated by colour. Each S-star is labelled, according to the Gillssen catalogue \citep{Gillessen_2009}. A colour gradient can be seen, ranging from orange in the bottom right to pink in the top left.}
        \label{fig:ecc_sma}
    \end{figure*}

    To map the separations of S-star curves and Sgr A*, we subtract the black hole curve from each S-star curve, and take the temporal mean for each S-star; we define this as $\left\langle \Delta^{\mathrm{S\text{-}star\vphantom{{}_\mathrm{BH}}}}_{\mathrm{BH}}\log_{10}(\delta_r)\right\rangle$. To find a relationship between this and the Keplerian elements of the stars, we use the initial orbital parameters from \citep{Gillessen_2009}. We expect to see that a low semi-major axis ($a$) corresponds to a higher magnitude in $\delta_r$, as stars that are close to the black hole should be more sensitive to the changes inside the system. Moreover, eccentricity ($e$) and $\delta_r$ should be directly correlated, since a highly elliptical orbit has a closer pericenter to the black hole. In \autoref{fig:ecc_sma}, we show the relationship between $a$, $e$ and $\left\langle \Delta^{\mathrm{S\text{-}star\vphantom{{}_\mathrm{BH}}}}_{\mathrm{BH}}\log_{10}(\delta_r)\right\rangle$.
    The pattern that emerges in the colour gradient in \autoref{fig:ecc_sma} hints that eccentricity and semi-major axis play a major role in the curve separations. The influence of eccentricity and semi-major axis on the mean separation are both important. The other orbital parameters can be compared with each other in a similar fashion, however, no apparent pattern is found. This observation suggests that their influence on $\left\langle \Delta^{\mathrm{S\text{-}star\vphantom{{}_\mathrm{BH}}}}_{\mathrm{BH}}\log_{10}(\delta_r)\right\rangle$ is negligible, compared to that of the eccentricity and semi-major axis. 
    Furthermore, the stars with the highest colour values in \autoref{fig:ecc_sma} are S21 and S29. \citep{portegies-zwart-2023} demonstrate that S21 is one of the stars that were part of the big event at 2876 yr. Moreover, S29 was shown to have 6 close encounters during the simulation. In addition, S67 has 10 close encounters, the most of any star. Therefore, despite its relatively low eccentricity and large semi-major axis, it is not green in \autoref{fig:ecc_sma} as opposed to its nearest neighbour, S87.
%-----------------------------------------------------------------
\section{Symbolic regression}
    We adopt PySR \citep{cranmer2023interpretable}, a symbolic regression Python package for discovering interpretable analytical equations that describe an underlying pattern in a dataset. Only semi-major axis and eccentricity are used in the model, since symbolic regression may not provide accurate results when the dimensionality of the data is high \citep{matchev2021analytical}. Moreover, PySR does not correctly interpret the evolution of $\delta_r$, $a$ and $e$ for each star, suggesting it does not support time-series data\footnote{Eureqa (not open-source) claims to be able to handle time-series data (see \url{https://www.creativemachineslab.com/eureqa.html}).}. Therefore, we provide the model with $\left\langle \Delta^{\mathrm{S\text{-}star\vphantom{{}_\mathrm{BH}}}}_{\mathrm{BH}}\log_{10}(\delta_r)\right\rangle$ and the initial semi-major axis and eccentricity. The results can be found in  \autoref{tab:symfinal}.
    \begin{table*}[h]
    \begin{center}
    \caption{Symbolic regression results on initial semi-major axis [au] and eccentricity and mean offsets between S-star and black hole.  Here, $y =\left\langle \Delta^{\mathrm{S\text{-}star\vphantom{{}_\mathrm{BH}}}}_{\mathrm{BH}}\log_{10}(\delta_r)\right\rangle$.   The optimal equation is of complexity 8. The loss parameter is defined by the least squares error, $L(r) = |r|^2$, where $r$ is the difference between the target and prediction variables \citep{cranmer2023interpretable}. The score compares the loss to complexity; our model seeks the lowest loss at the lowest complexity.}
    \begin{tabular}{@{}cccc@{}}
    \toprule
    Equation & Complexity & Loss & Score \\
    \midrule
    $y = 2.7747$ & $1$ & $1.9979$ & $0.0$ \\
    $y = 4.1060 e$ & $3$ & $0.75787$ & $0.48467$ \\
    $y = 4.0876 - 0.00023120 a$ & $5$ & $0.51127$ & $0.19681$ \\
    $y = e \left(5.1075 - 0.00023120 a\right)$ & $7$ & $0.25461$ & $0.34859$ \\
    $y = - 0.00018446 a + \ln{\left(e \right)} + 4.3103$ & $8$ & $0.15075$ & $0.52408$ \\
    \begin{minipage}{0.5\linewidth}
    \centering
    $y = - 0.00016205 a + e + \ln{\left(e \right)} + 3.5003$\end{minipage} & $10$ & $0.12189$ & $0.10626$ \\
    \begin{minipage}{0.5\linewidth}  
    \centering
    $y = - 0.00016753 a + e + 0.88260 \ln{\left(e \right)} + 3.4742$\end{minipage} & $12$ & $0.11897$ & $0.012141$ \\
    \bottomrule
    \end{tabular}
    \label{tab:symfinal}
    \end{center}
    \end{table*}
    
    From the optimal equation, which balances accuracy and simplicity, we can derive the following:
    \begin{equation}\label{eq:pysrfinal}
        y = - 0.00018446a/\mathrm{au}  + \ln(e) + 4.3103.
    \end{equation}
    $y$ is a measure of $\log_{10}(\delta_r)$,
    \begin{equation}
    \left\langle\log_{10}(\delta_r)\right\rangle = - 0.00018446 a/\mathrm{au}  + \ln(e) + 4.3103.
    \end{equation}
    We use this to estimate $\delta_r$ approximately as
    \begin{gather}
        \left\langle\delta_r\right\rangle = 2\cdot10^4 10^{-0.00018446a/\mathrm{au}} e^{2.3} \\
        \Rightarrow \left\langle\delta_r\right\rangle \propto e^{2.3}.
    \end{gather}
    
    Subtracting \autoref{eq:pysrfinal} from $\log_{10}(\delta_r)$ reduces the offsets significantly, as shown in the right panel of \autoref{fig:pysrfinal}.
%-----------------------------------------------------------------
\section{Conclusions}
We aimed to characterize the individual behaviour of stars during the evolution of a Newtonian chaotic dynamical system. Using a gravitational $N$-body simulation of the S-star cluster, we derive the Lyapunov timescale of the system over $10^4$ yr to be $\sim$420 yr. 

In the position space of the S-stars, we find a vertical offset between their curves. A relationship with the Keplerian orbits of the S-star was proposed to explain this offset in $\delta_r$. We adopt symbolic regression to find the relation $\left\langle\delta_r\right\rangle = 2\cdot10^4 10^{-0.00018446a/\mathrm{au}} e^{2.3}$, where $a$ and $e$ are taken from the initial orbit. We conclude that $\left\langle\delta_r\right\rangle \propto e^{2.3}$; the time-averaged individual phase-space distance with respect to the black hole increases rapidly with the star's orbital eccentricity.

\begin{acknowledgements}
    This publication is funded by the Dutch Research Council (NWO) with
    project number OCENW.GROOT.2019.044 of the research programme NWO XL. It
    is part of the project “Unravelling Neural Networks with
    Structure-Preserving Computing”. In addition, part of this publication
    is funded by the Nederlandse Onderzoekschool Voor Astronomie (NOVA). 
    
    TB’s research was supported by an appointment to the NASA Postdoctoral
    Program at the NASA Ames Research Center, administered by Oak Ridge
    Associated Universities under contract with NASA.
    
    We greatly thank the referee for taking the time to read this manuscript carefully and providing us with well-considered comments.
\end{acknowledgements}

% WARNING
%-------------------------------------------------------------------
% Please note that we have included the references to the file aa.dem in
% order to compile it, but we ask you to:
%
% - use BibTeX with the regular commands:
%   \bibliographystyle{aa} % style aa.bst
%   \bibliography{Yourfile} % your references Yourfile.bib
%
% - join the .bib files when you upload your source files
%-------------------------------------------------------------------

\bibliographystyle{aa} % style aa.bst
\bibliography{references}

\end{document}